\documentclass[twocolumn,amssymb,amsmath,floatfix,superscriptaddress,nofootinbib,aps,prc]{revtex4-2}

\usepackage{epsfig}
\usepackage{graphicx}
\usepackage{float}
\usepackage{hyperref}
\usepackage{amsmath}
\usepackage{amssymb}

\begin{document}
	
\title{ Momentum-dependent flow correlations in deformed nuclei at collision
   energies available at the BNL Relativistic Heavy Ion Collider}

\author{Rupam Samanta}
\email{rsamanta@agh.edu.pl}
\affiliation{AGH University of Science and Technology, Faculty of Physics and
	Applied Computer Science, aleja Mickiewicza 30, 30-059 Cracow, Poland}	

\author{Piotr Bo\.zek}
\affiliation{AGH University of Science and Technology, Faculty of Physics and
	Applied Computer Science, aleja Mickiewicza 30, 30-059 Cracow, Poland}

\begin{abstract}
  Flow fluctuations in ultra-relativistic heavy-ion collision can be probed by studying the momentum dependent correlations  or the factorization-breaking coefficients between  flow harmonics in separate kinematic bins (transverse momentum $p$ or pseudorapidity $\eta$). We study such factorization-breaking
  coefficients for collisions of deformed $^{238}$U+$^{238}$U nuclei to see the effect of the nuclear deformation on momentum dependent  coefficients. We also study momentum dependent mixed-flow correlations for the isobar collision system : $^{96}$Ru+$^{96}$Ru and $^{96}$Zr+$^{96}$Zr, which have the same mass number but different nuclear structure, thus providing the ideal scenario to study nuclear deformation effect on such observables. We use the  TRENTO + MUSIC  model for simulations and event-by-event analysis of those observables. We find that these momentum dependent correlation coefficients are not only excellent candidates to probe the fluctuation in heavy-ion collision, but also  show significant sensitivity to the nuclear deformation.  
\end{abstract}


\maketitle

\section{Introduction}\label{introduction}
In ultrarelativistic heavy-ion collisions the dense matter, known as  Quark-Gluon Plasma (QGP), is created in the interaction region.  One of the goals of theoretical and experimental studies in heavy-ion collision is to investigate the properties of the QGP medium. The evolution and expansion of the dense matter  can be modeled  through viscous hydrodynamics. Another interesting feature is to study the initial state in  heavy-ion collisions and how specific properties of the  initial state propagate to the final state of  the collision.  Through the viscous hydrodynamic evolution of the QGP medium, the spatial anisotropy in the source distribution at the initial state  gives rise to the  momentum anisotropy of the distribution of emitted  particles. This momentum anisotropy in azimuthal angle  is characterized by the Fourier expansion of the momentum distribution of the particles. The Fourier coefficients are called the \textit{harmonic flow coefficients} \cite{Huovinen:2006jp,Voloshin:2008dg,Hirano:2008hy,Heinz:2009xj,Heinz:2013th,Florkowski:2014yza}. 

The initial state in  heavy-ion collisions fluctuates event-by-event, as a result of which the final state flow harmonics also fluctuate event-by-event \cite{Aguiar:2001ac,Takahashi:2009na,Alver:2010gr,Schenke:2010rr,Schenke:2012wb}. These fluctuations cause a  decorrelation between the flow vectors in separate kinematic (transverse momentum $p$ or pseudorapidity $\eta$) bins \cite{Bozek:2010vz,Gardim:2012im,Kozlov:2014fqa,Gardim:2017ruc}. The decorrelation between two harmonic flow vectors in two transverse momentum bins denotes the deviation of the correlation coefficients from 1. This correlation coefficient  is called the \textit{factorization-breaking coefficient} \cite{Gardim:2012im}. The flow vector decorrelation involves both the flow magnitude and the flow angle decorrelation \cite{Jia:2014ysa}. There has been several recent theoretical \cite{Bozek:2018nne,Barbosa:2021ccw,Bozek:2021mov,Magdy:2022jai,Nielsen:2022jms} and experimental studies \cite{ATLAS:2017rij,ALICE:2017lyf,ALICE:2022smy} which have shown that the study of these momentum dependent correlations between harmonic flows or the factorization-breaking coefficients can nicely probe  event-by-event fluctuations in  heavy-ion collisions.  

In our previous work \cite{Bozek:2021mov,Bozek:2022slu}, we studied such factorization-breaking coefficients for collisions of spherical Pb+Pb nuclei  at the
Large Hadron Collider energies. In many cases, the atomic nuclei used in heavy-ion collision experiments are  not spherical but rather deformed, with   quadrupole or octupole deformation \cite{Giacalone:2019pca,Jia:2021tzt}. Such deformations affect the initial state of the heavy-ion collision and introduce additional fluctuations due to the random orientation of these deformed nuclei. The deformed structure of the colliding nuclei enhances the initial eccentricity which in turn enhances the final state flow harmonic vectors \cite{Giacalone:2018apa,Zhang:2021kxj}. In the past few years, there has been many theoretical  \cite{Giacalone:2018apa,Giacalone:2019pca,Giacalone:2020awm,Giacalone:2021udy,Giacalone:2021uhj,Jia:2021oyt,Jia:2021qyu,Jia:2021tzt,Jia:2021wbq,Jia:2022iji,Jia:2022qgl,Jia:2022qrq,Zhang:2021kxj,Zhang:2022fou} and some experimental studies \cite{PHENIX:2015tbb,STAR:2015mki,STAR:2021mii} on   how the nuclear deformation affects the final state flow harmonics and flow correlation coefficients. It would be interesting to study the factorization-breaking coefficients for the collisions of  deformed nuclei, e.g. $^{238}$U+$^{238}$U collision at energies of the
BNL  Relativistic Heavy Ion Collider (RHIC). In particular, it may be worth to investigate  how the decorrelation between the flow vectors, flow magnitudes, and flow angles is affected by the deformed shape of the colliding nuclei. 

Moreover, recent theoretical \cite{Giacalone:2021uhj,Jia:2022iji,Jia:2022qgl,Jia:2022qrq,Zhang:2021kxj,Zhang:2022fou} and experimental studies \cite{STAR:2021mii} on the isobar collisions have drawn the  attention of the community to studies of nuclear deformation
in heavy-ion collisions. Isobars which have equal mass but different nuclear deformation  are the ideal candidate to probe the nuclear deformation effect in heavy-ion collision, e.g $^{96}$Ru+$^{96}$Ru  and $^{96}$Zr+$^{96}$Zr collision at the top  RHIC energy. Any difference in the final state observables between them is due to the difference in their nuclear structure. For the isobar collisions, the non linear mixed flow correlations, e.g. $V_2^2-V_4$, $V_2V_3-V_5$ correlations, are more sensitive to the  nuclear deformation than the factorization-breaking coefficients. In \cite{Jia:2022qrq} such nonlinear correlations between the global (momentum averaged) flow vectors were studied. In this paper, we look at the momentum dependent, nonlinear correlations between flow harmonics of different order.

We study in the  hydrodynamic model  different  factorization-breaking coefficients for collisions of deformed U+U, Zr+Zr, Ru+Ru  nuclei and compare the results with collisions of  spherical nuclei. Next, we study the momentum dependent nonlinear mixed flow correlations for  U+U collisions and for the isobar collisions Ru+Ru and Zr+Zr. We compare the results for the factorization-breaking coefficients and nonlinear harmonic flow correlations  for  central ( 0-5\% ) and for ultra-central ( 0-1\% ) collisions.

\section{Momentum dependent flow correlation : Factorization-breaking coefficients}
\label{factbreaktheory}

The azimuthal anisotropy in the distribution of hadrons emitted in a
heavy-ion collision is described by the Fourier expansion,
\begin{equation}
\frac{d^2N}{dp d\phi} = \frac{dN}{2\pi dp} \left(  1 + 2 \sum_{n=1}^{\infty} V_n(p)e^{i n \phi} \right) \ \ ,
\label{ptdist}
\end{equation}
where 
 $V_n(p)$ is the flow vector for the $n^{th}$ order harmonic flow, which depends on the particle  transverse momentum $p$  ( usually denoted by $p_t$ or $p_T$ ) and  is calculated in the mid-rapidity region. The flow vector is further written as $V_n(p)$ = $v_n(p)e^{i n \Psi_n (p)}$, where $v_n(p)$ and $\Psi_n (p)$ are the corresponding momentum dependent flow magnitude and flow angle  respectively. The global or the momentum averaged harmonic flow vector in an event is calculated by integrating the momentum dependent flow $V_n(p)$ over the whole momentum range with respect to  the momentum distribution (Eq. \ref{ptdist}) as
\begin{equation}
	V_n = \frac{\int_{p_{min}}^{p_{max}} dp V_n(p)\frac{dN}{dp d\phi}}{\int_{p_{min}}^{p_{max}} dp \frac{dN}{dp d\phi}} \ ,
	\label{globalflow}
\end{equation}  
where the denominator represents the total charged particle multiplicity $N$. The average of the flow harmonics over a sample of events in a given centrality bin is calculated through the two-particle cumulant method as
\begin{equation}
	v_n\{2\} = \sqrt{\langle V_n V_n^* \rangle } \ ,
\end{equation} 
where $\langle \dots \rangle$ denotes the event-average. 
For the momentum-dependent flow the above equation takes the form
 \begin{equation}
 	v_n\{2\}(p) = \frac{\langle V_n V_n^*(p) \rangle }{\sqrt{\langle V_n V_n^* \rangle }} \ .
 \end{equation} 


 One  way to probe  flow fluctuations is to study the correlations between flow harmonics in different kinematic (transverse momentum or pseudorapidity) bins \cite{Bozek:2010vz,Gardim:2012im}.  Event by event fluctuations of the flow harmonics cause a significant decorrelation (deviation from 1 of the correlation coefficient) between the flow vectors in two different momentum bins. 
To construct such a correlation coefficient, two flow vectors are taken from two momentum bins $p_1$ and $p_2$ and the correlation coefficient between them is defined as
\begin{equation}
r_n(p_1,p_2) = \frac{\langle V_n(p_1)V_n^*(p_2)\rangle }{\sqrt{\langle v_n^2(p_1)\rangle \langle v_n^2(p_2)\rangle}}  \ .
\label{veccorr}
\end{equation}  
Such a correlation coefficient is called the \textit{factorization-breaking coefficient} and it is a two-particle correlator. The coefficients $r_n(p_1,p_2)$ have been measured in experiments \cite{CMS:2013bza,Zhou:2014bba,CMS:2015xmx} and studied in models \cite{Gardim:2012im,Kozlov:2014fqa,Gardim:2017ruc,Bozek:2018nne,Barbosa:2021ccw}, where significant decorrelation has been observed between the two flow vectors at two different momenta. However, the flow vector decorrelation is a combined effect of the flow magnitude and the flow angle decorrelation \cite{Jia:2014ysa,Heinz:2013bua}. Both have been found to contribute almost equally to the total flow vector decorrelation \cite{ATLAS:2017rij}. Similar factorization coefficients as in Eq. (\ref{veccorr}) could be constructed between the flow magnitudes in two momentum bins and flow angle decorrelation could be calculated from the ratio of the flow vector and flow magnitude factorization coefficients.

While formally possible in many models, such correlations coefficients between the flow magnitudes cannot be measured in experiments.
To measure the flow magnitude and flow angle decorrelations, such
factorization-breaking coefficients need to be constructed in the second moment, between the squares of the flow vectors \cite{Jia:2017kdq,Bozek:2018nne}. Similar to Eq. (\ref{veccorr}) one can construct factorization-breaking coefficients between the squares of flow vectors  and the squares of the flow magnitudes, both of which could be measured experimentally. From the ratio of these two factorization-breaking coefficients the  flow angle decorrelation could be calculated. However, such correlation coefficients  involve four-particle correlators, which are difficult to measure in experiment due to limited statistics in  high transverse momentum bins. 

 To ease this difficulty, similar factorization-breaking coefficients can be constructed keeping one of the flow harmonics in a small  momentum bin and the other flow harmonic as momentum averaged (global flow). Such correlation coefficients, where only one bin in transverse momentum is used, are experimentally accessible  \cite{Heinz:2013bua,ALICE:2017lyf,ALICE:2022smy,ALICE:2017lyf,Magdy:2022jai}.
 Such factorization-breaking coefficients for higher moments of the  harmonic flow can be constructed as \cite{Nielsen:2022jms,Bozek:2021mov}
\begin{equation}
  	r_{n;2}(p) = \frac{\langle V_n^2V_n^*(p)^2\rangle }{\sqrt{\langle v_n^4\rangle \langle v_n^4(p)\rangle}}
  	\label{vecsqcorr}
\end{equation}      
for the flow vector square and
\begin{equation}
\begin{aligned}
	r_n^{v_n^2}(p) &=  \frac{\langle | V_n^2 | |V_n^*(p)^2|\rangle }{\sqrt{\langle v_n^4\rangle \langle v_n^4(p)\rangle}} \\
	&=  \frac{\langle  v_n^2  v_n^2(p)\rangle }{\sqrt{\langle v_n^4\rangle \langle v_n^4(p)\rangle}} 
	\label{magsqcorr}
\end{aligned}
\end{equation}
 for the flow magnitude square. With the flow vector and flow magnitude squared factorization coefficients from Eq. (\ref{vecsqcorr}) and Eq. (\ref{magsqcorr}) respectively, the flow angle decorrelation can be estimated from the ratio of the two as,
 \begin{equation}
 	\begin{aligned}
 	F_n(p) &= \frac{\langle V_n^2V_n^*(p)^2\rangle  }{\langle  v_n^2  v_n^2(p)\rangle } \\
 	&= \frac{\langle  v_n^2  v_n^2(p)\cos[2n(\Psi_n-\Psi_n(p))]\rangle }{\langle  v_n^2  v_n^2(p)\rangle}  \ .
 	\label{angcorr}
\end{aligned}
\end{equation}
The above formula could be used in experiment to estimate the flow angle decorrelation. In practice the momentum dependence of the flow magnitudes in Eq. (\ref{angcorr}) can be dropped and the flow angle decorrelation $F_n(p)$  is an excellent approximation of the flow angle decorrelation \cite{Bozek:2021mov}
 \begin{equation}
 \frac{\langle  v_n^4 \cos[2n(\Psi_n-\Psi_n(p))]\rangle }{\langle  v_n^4 \rangle} 
 \label{wtdangcorr} \ .
\end{equation}
The above formula represents the average of the flow angle correlation weighted by some power of the flow magnitude, whereas the simple average $\langle  \cos[2n(\Psi_n-\Psi_n(p))]\rangle$ gives a very different result \cite{Bozek:2022slu}.

In \cite{Bozek:2021mov,ALICE:2022smy} such factorization-breaking coefficients have been studied  for  Pb+Pb collisions. It would be interesting to study such coefficients for central  collisions of deformed  nuclei e.g. for U+U collisions or for the isobaric collisions (Ru+Ru and Zr+Zr) at RHIC energies. Such studies could reveal interesting effect of the deformation of the colliding nuclei on these correlation coefficients as well as some new information on their structure. 

\section{Collision of deformed nuclei}\label{deformednuclei}

In the past few years there have been many studies on collective flow in  relativistic  collision of deformed nuclei  in experiments \cite{PHENIX:2015tbb,STAR:2015mki,STAR:2021mii}  and quite comprehensive studies in  models \cite{Giacalone:2018apa,Giacalone:2019pca,Giacalone:2020awm,Giacalone:2021udy,Giacalone:2021uhj,Jia:2021oyt,Jia:2021qyu,Jia:2021tzt,Jia:2021wbq,Jia:2022iji,Jia:2022qgl,Jia:2022qrq,Zhang:2021kxj,Zhang:2022fou}. These studies have disclosed many interesting consequences of nuclear structure of such nuclei in high energy regime and  such studies in heavy ion collision could  be  phenomenologically very important. The shape of the atomic nucleus is commonly modeled through the Woods-Saxon density distribution\cite{Giacalone:2019pca,Giacalone:2021uhj,Jia:2022qrq,Zhang:2022fou} in the form
\begin{equation}
  \rho(r, \theta, \phi ) = \frac{\rho_0}{1+e^{\frac{[r-R_0(1+\beta_2 Y^0_2(\theta, \phi)+\beta_3 Y^0_3(\theta, \phi))]}{a_0}}} \ ,
  \label{eq:WS}
\end{equation}
where, $\rho_0$ is the maximum nuclear matter density and $Y^m_l$ are the spherical harmonics. The four structure parameters $a_0$, $R_0$, $\beta_2$ and $\beta_3$ represent the nuclear diffusivity, half-width radius, quadrupole deformation, and octupole deformation respectively.  

Past studies have shown that the initial state in  heavy-ion collisions depends on the deformation of the colliding nuclei. At the initial state of the collision, the participant nucleons deposit energy (or entropy) in the overlap region, forming a hot dense QGP  fireball. The shape of this overlap region is sensitive to the nuclear deformation. In general, the geometry of this initial overlap region is characterized by its eccentricity,
\begin{equation}
\epsilon_n e^{i\Phi_n} = - \frac{\int r^n s(r,\phi) e^{i n \phi} r dr d\phi}{\int r^n s(r,\phi)  r dr d\phi}   \ ,
\end{equation}
where $s(r,\phi)$ denotes the entropy deposited in the overlap region. If the colliding nuclei are deformed, they have a non-zero $\beta_n$ which enhances the magnitude of the eccentricity $\epsilon_n$, which in general follow the simple relation\cite{Jia:2021tzt} for n=2 and 3,
\begin{equation}
\epsilon_n\{2\}^2 = a'_n + b'_n \beta_n^2+\sum_{m \neq n}b'_{n,m} \beta_m^2 \ .
\label{epsilonbeta}
\end{equation} 
Through the hydrodynamic evolution of the initial fireball, the eccentricity $\epsilon_n$ is translated into the azimuthal momentum anisotropy of the final state hadrons. The harmonic flow coefficients $v_n$ follow the phenomenological relation for n=2 and 3~:
\begin{equation}
  	v_n\{2\}^2 = k_n  \epsilon_n\{2\}^2   \ .
\end{equation}  
As a result, the harmonic flow depends on the nuclear deformation through a similar parametric dependence as Eq. (\ref{epsilonbeta}) for the elliptic (n=2) and triangular flow (n=3), 
\begin{equation}
	v_n\{2\}^2 = a_n + b_n \beta_n^2+\sum_{m \neq n}b_{n,m} \beta_m^2   \ .
	\label{vnbeta}
\end{equation}
The harmonic flow is dominated by the elliptic flow $v_2$ and the triangular flow $v_3$ and these two harmonics are primarily affected by the deformation of the colliding nuclei.
The relative importance of the deformation on the initial eccentricities is especially pronounced for central collisions.
In central collision of spherical nuclei, both harmonic flows $v_2$ and $v_3$ are primarily determined by  fluctuations of the initial state  and are small.
The additional increase of the initial eccentricity for collisions of deformed nuclei is significant and could be indirectly measured in heavy-ion collision  experiments.

\begin{table}
	\centering
\begin{tabular}{|c|c|c|c|c|}
 \hline
  Species  & $R_0$ (fm) &  $a_0$  (fm) &  $\beta_2$  &  $\beta_3$  \\
 \hline
 $^{238}$U & 6.86 & 0.42 & 0.265 & 0\\
 \hline
 $^{96}$Ru & 5.09 & 0.46 & 0.162 & 0\\
 \hline
 $^{96}$Zr & 5.02 & 0.52 & 0.06 & 0.20\\
 \hline
\end{tabular} 
\caption{Parameters of the Woods-Saxon density distribution  (\ref{eq:WS}) for  $^{238}$U,  $^{96}$Ru and $^{96}$Zr nuclei.}
\label{wsparamtable}
\end{table}

We perform event by event simulations of U+U collisions at  $\sqrt{s_{NN}}$=193~GeV and for  isobar  Ru+Ru and Zr+Zr collisions at $\sqrt{s_{NN}}$=200~GeV.  The initial density in the fireball is  taken from the  parametric TRENTO model \cite{Moreland:2014oya}. For the hydrodynamic evolution we use the boost invariant 2+1-dimensional  MUSIC viscous hydrodynamic code \cite{Schenke:2010nt,Schenke:2010rr,Paquet:2015lta}. The  shear viscosity to entropy density ratio is ${\eta}/{s}$ = 0.12. At the freeze-out temperature of $135$MeV a Cooper-Frye emission of hadrons is performed, with resonance decays. The emission spectra (in transverse momentum and azimuthal angle)  are calculated for each hydrodynamic event. From the spectra the flow observables at midrapidity (flow harmonics) are obtained by integration. 
The equation of state used is s95p \cite{Huovinen:2009yb},
with a chemical freeze-out at the temperature of $155$MeV.

The Wood-Saxon density parameters, including the deformation parameters,  are presented in Tab. \ref{wsparamtable}. We choose a parametrization with an quadrupole deformation only for U and Ru,  but the nuclear shape for Zr has small quadrupole and a sizable octupole deformation. The Uranium nucleus could have a small octupole deformation besides the dominant quadrupole one. In this paper we study only the leading effect and  we chose a parametrization without octupole deformation for U. We have simulated in total 7000 U+U collision and 20000 Ru+Ru and  Zr+Zr minimum-bias events. In most cases,  5000 events would provide  sufficient statistics, but for the isobar analysis a larger number of events is required to increase the precision for the comparative studies between the isobar collision Ru+Ru and Zr+Zr, as well as to study the ratio of the observables between Ru and Zr (Sec. \ref{sec:ratioandcent}). The centrality selections are made according to the total entropy  deposited at the overlap region in the initial state, which is proportional to the integrated initial density in the transverse plane obtained  in TRENTO model with the parameter $p=0$ \cite{Moreland:2014oya}.

The harmonic flow coefficients $v_n$  are calculated  for all charged particles with transverse momenta between $0.2$
and $3.0$GeV.  The momentum dependent correlators are calculated by averaging over events in the given centrality class. In the following two subsections we present the results for the factorization-breaking coefficients  given by Eqs. \ref{vecsqcorr}, \ref{magsqcorr}, and \ref{angcorr} for collisions of deformed nuclei  and compare them to the corresponding spherical nuclei collision results. Comparing to the spherical nuclei with $\beta_n$ = 0, one can isolate  the effect of nuclear deformations on these correlations.

\subsection{Results for U+U collisions at $\sqrt{s_{NN}}=$ 193 GeV}

\label{sec:UUresults}

In this section, we present the results for the factorization-breaking coefficients for flow vector square and flow magnitude square (Eqs.  \ref{vecsqcorr} and \ref{magsqcorr}), as well as for the flow angle decorrelation (Eq. \ref{angcorr}).   $^{238}$U nuclei are deformed with a  leading quadrupole deformation characterized by $\beta_2 > 0$, leading to a a {\it prolate} structure \cite{Jia:2021tzt}.
In each collision, the two colliding nuclei  have random orientations of their deformation axes. The two extreme cases would be the tip-to-tip and the body-to-body collision  \cite{Giacalone:2019pca}. In the first case, the overlap region is circular and in the later one the overlap region has an elliptic shape, leading to a deformed initial state even in very central collisions. This anisotropic initial state ellipticity $\epsilon_2$ generates in  the final state a sizable elliptic flow $v_2$. We present results for collisions of  deformed U nuclei with $\beta=0.265$ and  for the corresponding spherical nuclei collision with $\beta= 0$, while keeping all other parameters same.

We consider two scenarios for the fluctuations in the initial deposited entropy.
In the default case, we use the exponential distribution for the distribution
of the entropy deposited by each participant nucleon \cite{Moreland:2014oya}.
The fluctuations in the entropy distribution partly wash out the
geometrical effect of the deformation in central collisions. For comparison, a calculation without additional fluctuations in the deposition of each participant nucleon is also considered.
 
\begin{figure}
	\vspace{5mm}
	\begin{center}
		\includegraphics[width=0.45\textwidth]{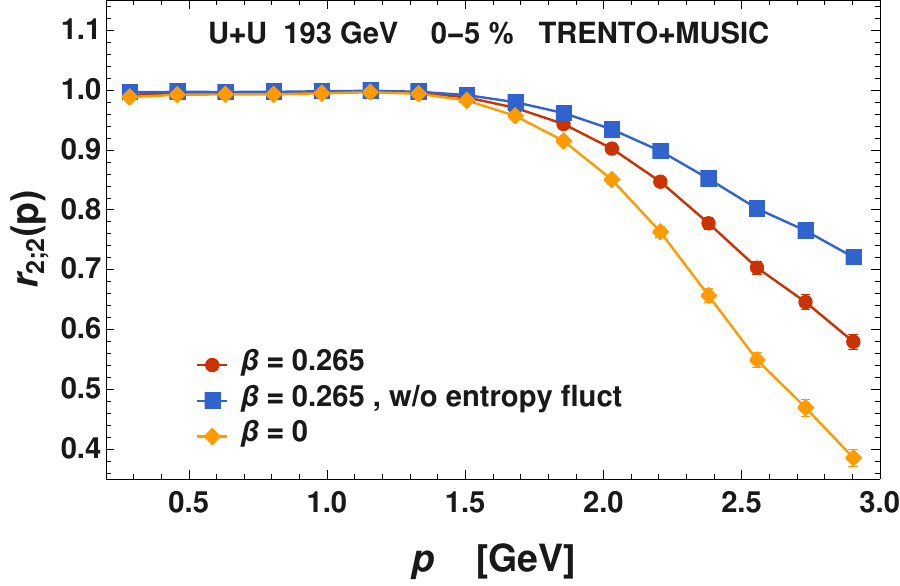} 
	\end{center}
	\caption{Flow vector factorization-breaking coefficient between $V_2^2$ and $V_2^2(p)$ as a function of transverse momentum $p$, for U + U collision at $\sqrt{s_{NN}}=$ 193 GeV for 0-5 \% centrality. The results with and without fluctuation in the initial entropy deposition are shown by the red dots and blue squares respectively. The orange diamonds represent the results for collisions of  spherical uranium nuclei.}
	\label{V2sqv2sqvecUU}
\end{figure}

In Fig. \ref{V2sqv2sqvecUU} the flow vector square factorization coefficients $r_{2;2}(p)$ for the elliptic flow are shown for  U+U collisions with 0-5\% centrality. The  correlation coefficient gradually deviates from 1 with  increasing transverse momentum, showing a significant decorrelation for both deformed (red dots) and spherical (orange diamonds) nuclei collision.  For collisions of spherical nuclei the elliptic flow is smaller and it easier to decorrelate the flow in different momentum bins, especially at higher momenta. When the fluctuations in the initial entropy deposition are switched off (blue squares), the decorrelation is smaller.

The presence of nuclear deformation increases the geometrical component in the
total eccentricity. This is analogous to the  case of semiperipheral collisions (at a relatively large impact parameter) \cite{Giacalone:2021uhj}. The overall magnitude of the flow is larger and the flow vectors in each  transverse momentum bin  are also more correlated with the overall orientation in the transverse plane. Hence the decorrelation between different momentum bins is small.

\begin{figure}
	\vspace{5mm}
	\begin{center}
		\includegraphics[width=0.45\textwidth]{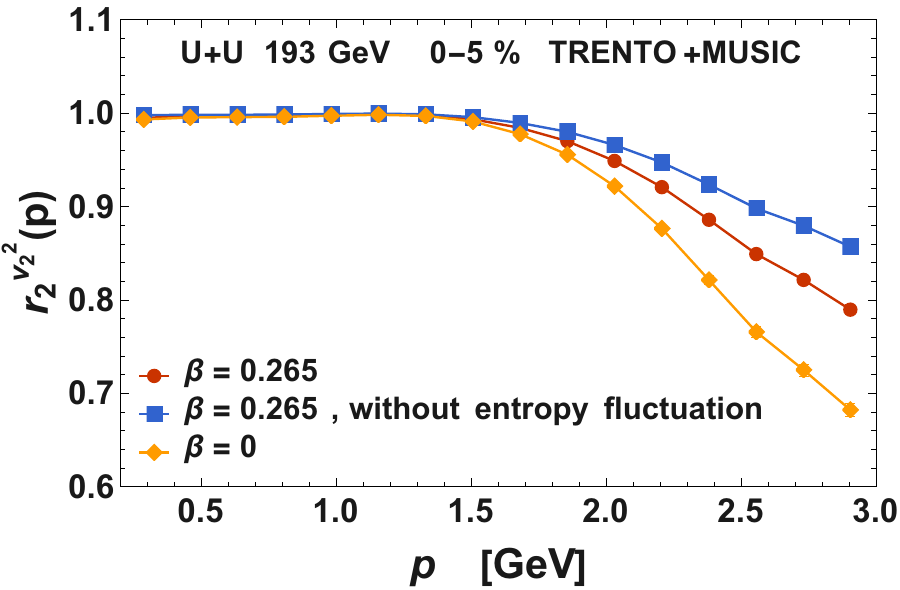} 
	\end{center}
	\caption{Flow magnitude square factorization-breaking coefficient between $v_2^2$ and $v_2^2(p)$ as a function of transverse momentum $p$, for U + U collision at $\sqrt{s_{NN}}=$ 193 GeV for 0-5 \% centrality. The symbols are similar as in  Fig.  \ref{V2sqv2sqvecUU}.}
	\label{V2sqv2sqmagUU}
\end{figure}

In Fig. \ref{V2sqv2sqmagUU} similar factorization-breaking  coefficients between the flow magnitude squares are shown for the elliptic flow. A similar trend  is observed for the three scenarios for collisions of deformed nuclei as for the flow vector decorrelation (see in Fig.  \ref{V2sqv2sqvecUU}). One of the interesting features to observe here is that the flow magnitude decorrelation  is approximately one half of the flow vector decorrelation
\begin{equation}
\left[1- r_{n;2}(p)\right] \simeq 2 \left[\ 1- r _{n}^{v_n^2}(p) \ \right] \ ,
\end{equation}
as noticed previously in related experimental and model studies
\cite{ATLAS:2017rij,Bozek:2021mov}.

\begin{figure}
	\vspace{5mm}
	\begin{center}
		\includegraphics[width=0.45\textwidth]{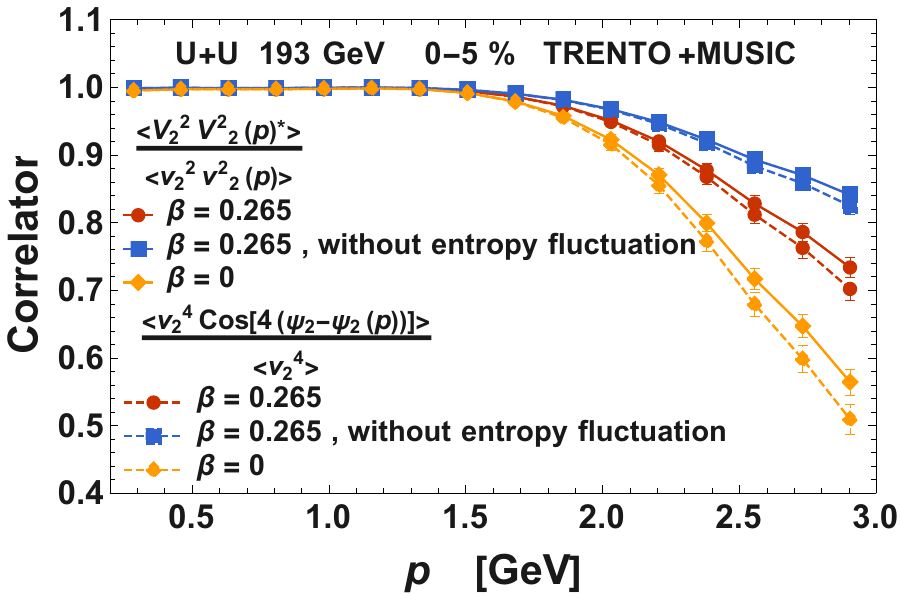} 
	\end{center}
	\caption{Flow angle decorrelation estimated from the ratio of the flow vector and the flow magnitude factorization-breaking coefficients for the elliptic flow for U + U collision at $\sqrt{s_{NN}}=$ 193 GeV for 0-5 \% centrality. Symbols with solid lines are the same as in Fig. \ref{V2sqv2sqvecUU}. The  dashed lines represent the corresponding magnitude weighted flow angle decorrelations calculated using Eq. \ref{wtdangcorr}.}
	\label{V2sqv2sqangUU}
\end{figure}

In Fig. \ref{V2sqv2sqangUU} the flow angle decorrelations are shown, calculated from the flow vector square and flow magnitude square factorization coefficients. The angle decorrelations calculated using Eq. \ref{angcorr} and Eq. \ref{wtdangcorr} are shown by the solid  and dashed line respectively. The two formulas give  similar results.  This shows that the experimentally accessible  estimate of angle decorrelation  (Eq. \ref{angcorr}) is a good approximation of (Eq. \ref{wtdangcorr}). 

\subsection{ Results for Isobaric collision : Ru+Ru and Zr+Zr at $\sqrt{s_{NN}}=$ 200 GeV}\label{sec:isobar}

\begin{figure}
	\vspace{5mm}
	\begin{center}
		\includegraphics[width=0.45\textwidth]{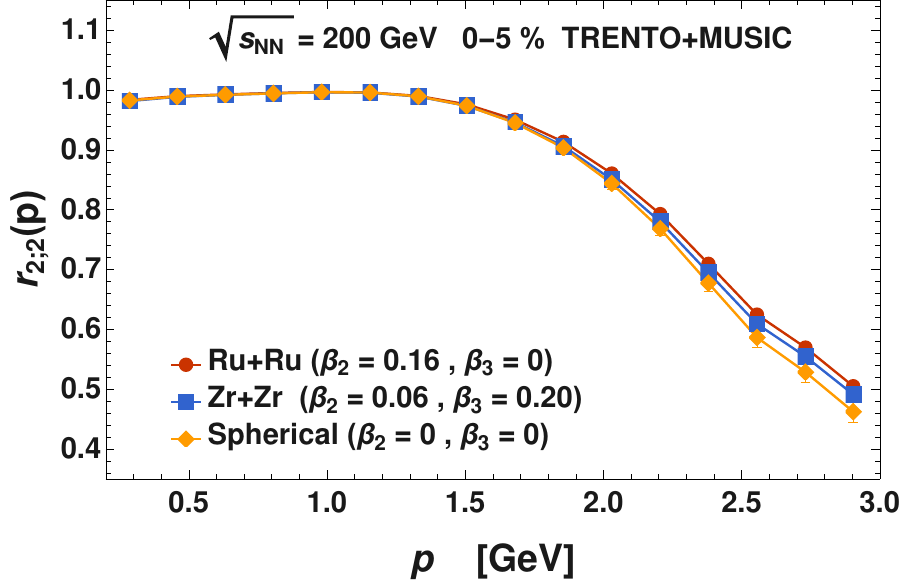} 
	\end{center}
	\caption{Flow vector factorization-breaking coefficient between $V_2^2$ and $V_2^2(p)$ as a function of transverse momentum $p$, for isobar collision at $\sqrt{s_{NN}}=$ 200 GeV for 0-5 \% centrality. The results for Ru+Ru and Zr+Zr  are shown by the red dots and blue squares respectively. The orange diamonds represent the results for the corresponding spherical Ru nuclei collision.}
	\label{V2sqv2sqvecRuZr}
\end{figure}

In this section, we present the results for the factorization-breaking coefficients  for isobaric collision systems e.g. $^{96}$Ru + $ ^{96}$Ru and $^{96}$Zr + $ ^{96}$Zr. Studies of such isobar collisions are a novel yet very useful probe of the deformation effect in high energy heavy ion collision, because the colliding nuclei have the same mass number but different nuclear deformation. Thus they are the ideal candidates for deformation studies as they can be easily studied in experiment. Any difference on the final state observables  between these two nuclei is due to the impact of the difference in their nuclear structures. However, not all observables have a significant sensitivity to this structural difference.  $^{96}$Ru has a large quadrupole deformation ($\beta_2$), whereas $ ^{96}$Zr  has a significant leading octupole deformation ($\beta_3$) as well, which make these isobars structurally different. The values of these deformation parameters are given in Tab. \ref{wsparamtable}. Past studies have established that $\beta_2$ and $\beta_3$ contribute to the initial eccentricity $\epsilon_2$ but only $\beta_3$ contributes to $\epsilon_3$. As a result, the final elliptic and triangular flow for these isobar collision are given by \cite{Jia:2021tzt,Zhang:2021kxj} 
\begin{equation}
 	v_2\{2\}^2 = a_2 + b_2 \beta_2^2 + b_{2,3} \beta_3^2 ,\ \  v_3\{2\}^2 = a_3 + b_3 \beta_3^2  \  . 
 \label{v2v3betaisobar}
\end{equation}
As we are focused on the ultra-central collisions, the 
In Fig. \ref{V2sqv2sqvecRuZr}, the flow vector square factorization-breaking coefficients for the elliptic flow $r_{2;2}(p)$ are shown for  0-5\% centrality for the isobar collisions.  The figure shows that there are significant decorrelations in both cases but there is almost no difference between Ru and Zr though they have different deformed structures. The reason behind this is that both  nuclei are deformed and both deformation parameters $\beta_2$ and $\beta_3$ contribute to $v_2$, although the type of deformations are different. Due to a small mass number, the fluctuation contribution to $\epsilon_n$ is relatively more important. The magnitude and angle decorrelations  show similar results (not shown). Thus the factorization-breaking coefficients for the elliptic flow are not suitable observables to study in isobar analysis.

\begin{figure}
	\vspace{5mm}
	\begin{center}
		\includegraphics[width=0.45\textwidth]{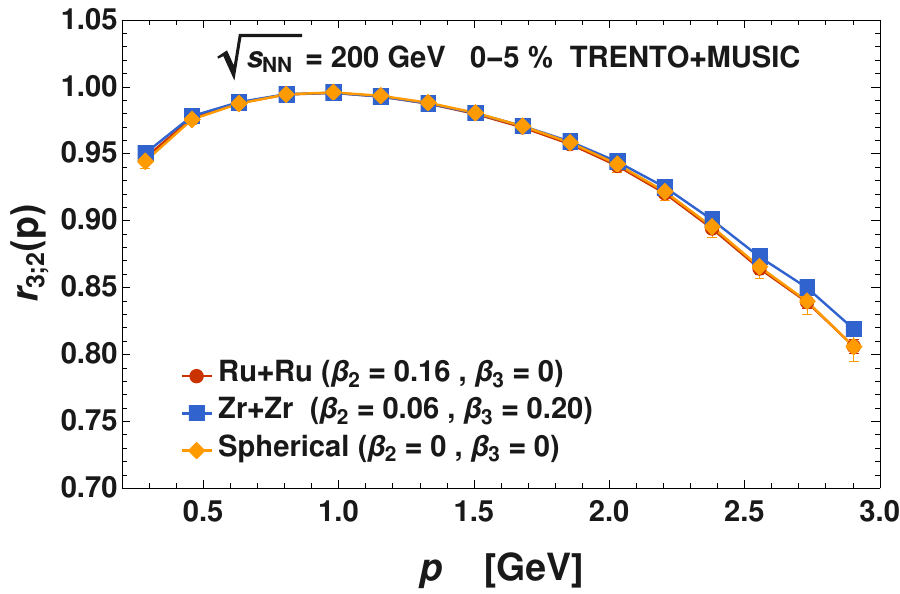} 
	\end{center}
	\caption{Flow vector factorization-breaking coefficient between $V_3^2$ and $V_3^2(p)$ as a function of transverse momentum $p$, for isobar collision at $\sqrt{s_{NN}}=$ 200 GeV for 0-5 \% centrality. Similar symbols as in Fig. \ref{V2sqv2sqvecRuZr}.}
	\label{fig:iso_r32}
\end{figure}

\begin{figure}
	\vspace{5mm}
	\begin{center}
		\includegraphics[width=0.45\textwidth]{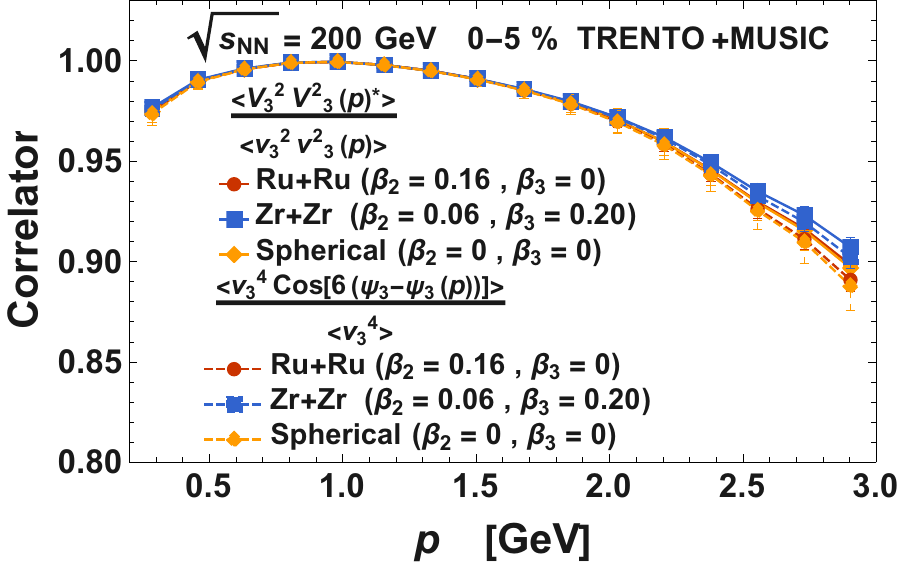} 
	\end{center}
	\caption{Flow angle correlations estimated from the ratio of flow vector and flow magnitude factorization coefficients for elliptic flow for isobar  collision at $\sqrt{s_{NN}}=$200 GeV for 0-5 \% centrality. Symbols with solid lines are the same as in Fig. \ref{V2sqv2sqvecRuZr}  . The  dashed lines represent the corresponding magnitude weighted flow angle decorrelations calculated using Eq. \ref{wtdangcorr}.}
	\label{fig:iso_r32ang}
\end{figure}

The analogous factorization-breaking coefficient for the square of the triangular flow $r_{3;2}(p)$ is shown in Fig. \ref{fig:iso_r32}. The results show  an effect of the octupole deformation of the colliding nuclei. The factorization-breaking coefficient is closer to 1 for Zr+Zr collisions with an octupole deformation. The results for collisions of spherical nuclei and nuclei with only quadrupole deformation are similar. Qualitatively similar results are visible for the angle decorrelations for collisions of nuclei with or without octupole deformation (Fig. \ref{fig:iso_r32ang}) or for the factorization-breaking coefficient for the magnitude of the triangular flow (not shown). However, quantitatively the effect is small and with the statistics used we cannot draw firm conclusions.

To check whether the deformation effect is consistent with the U+U collision results as compared to spherical nuclei, we also studied spherical Ru+Ru collision where $\beta_2$ and $\beta_3$ both are set to zero. In Fig. \ref{V2sqv2sqvecRuZr}, the results for the spherical Ru+Ru collision are denoted by the orange diamonds. The impact of  the deformation on the factorization-breaking coefficients for  Ru+Ru collisions is much smaller in comparison to U+U results. This is partly because of the fact that the deformation parameter $\beta_2$ is smaller in case of Ru , the ratio of the quadrupole deformation parameter, $ \beta^2_{2,{\text U}}/\beta^2_{2,{\text Ru}} = 2.676$. The other part consists of the different contribution to the event-by-event fluctuation depending on the system size and number of participants, when switching from deformed to spherical nuclei.   

 \section{Momentum dependent mixed flow correlations}\label{mixedflowcorr}
 


 The mixed-flow correlations, which are basically momentum dependent event-plane correlations, could serve as a measure of the nonlinear response in the hydrodynamic evolution of the QGP medium. They have been found to be sensitive to the deformation  of the colliding nuclei \cite{Jia:2022qrq}. Here we focus on two most important nonlinear correlators, $V_2^2-V_4(p)$ and $V_2V_3-V_5(p)$ correlations. In particular, we study the momentum dependent correlations
\begin{equation}
 \frac{\langle V_2^2V_4^*(p)\rangle }{\sqrt{\langle v_2^4\rangle \langle v_4^2(p)\rangle}} \ \ \text{and} \ \ \frac{\langle V_2V_3V_5^*(p)\rangle }{\sqrt{\langle v_2^2 v_3^2\rangle \langle v_5^2(p)\rangle}} \ ,
\label{mixedveccorr}
\end{equation} 
 for both isobar and uranium collisions. 

\begin{figure}
	\vspace{5mm}
	\begin{center}
		\includegraphics[width=0.47\textwidth]{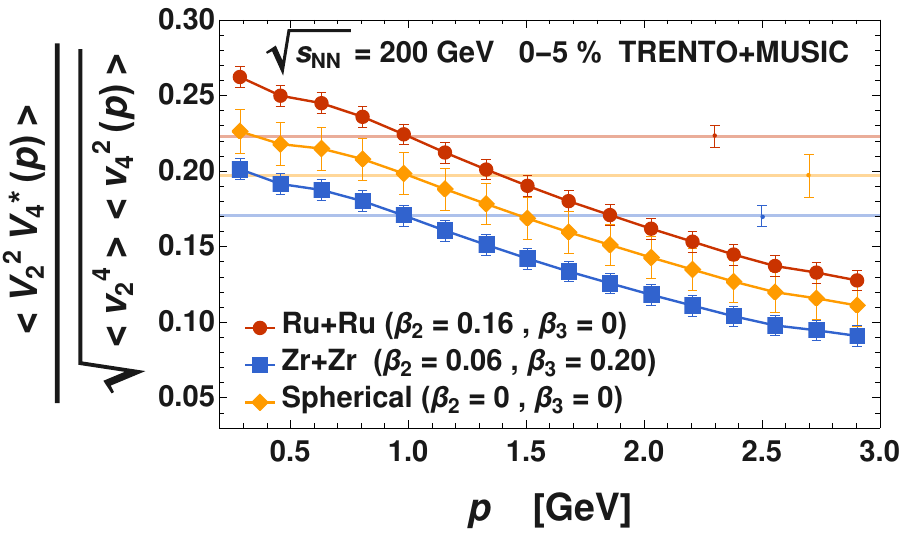} 
	\end{center}
	\caption{Nonlinear flow vector correlation between $V_2^2$ and $V_4(p)$ as a function of the  transverse momentum $p$, for isobar collision at $\sqrt{s_{NN}}=$ 200 GeV for 0-5 \% centrality. The results for Ru+Ru and Zr+Zr  are shown by the red dots and blue squares respectively. The orange diamonds represent the results for the corresponding spherical Ru nuclei collision. The horizontal lines represent the correlations between the momentum averaged flow vectors.}
	\label{V2sqv4vecRuZr}
\end{figure}

In Fig. \ref{V2sqv4vecRuZr}, the nonlinear flow vector correlations between $V_2^2$ and $V_4(p)$ are shown. The correlation coefficient is significantly larger for Ru+Ru than for Zr+Zr  collisions,  which  reflects the  difference in shape between the two nuclei having all other parameters similar. The nonlinear mixing between the two flow harmonics involved, $V_2$ and $V_4$ is sensitive to the deformation parameters. The difference is also obvious from the correlations between the momentum averaged flow (the horizontal  baselines of the plots).
It is interesting to notice the strong momentum dependence of the decorrelation.
 To have a complete picture, we have also calculated the mixed flow correlation coefficient for the corresponding spherical nuclei (orange diamonds in Fig. \ref{V2sqv4vecRuZr}), which surprisingly lie between the Ru and Zr. For Zr-Zr   collisions (with quadrupole and octupole deformation) the fourth order eccentricity $\epsilon_4$ is the largest, which leads to a decrease of the relative nonlinear contribution and reduces the value of the correlation coefficient between $V_2^2$ and $V_4$ for  that case. 

\begin{figure}
	\vspace{5mm}
	\begin{center}
		\includegraphics[width=0.47\textwidth]{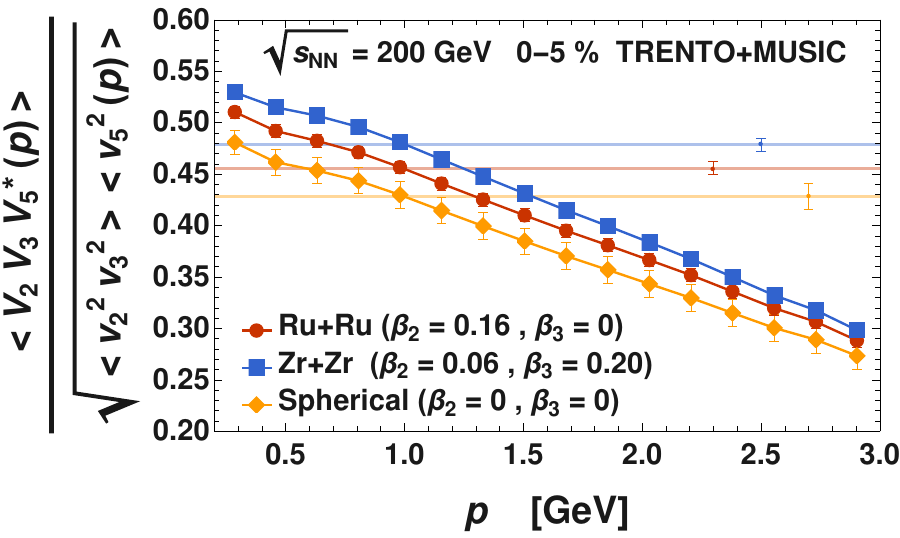} 
	\end{center}
	\caption{Nonlinear flow vector correlation between $V_2V_3$ and $V_5(p)$ as a function of transverse momentum $p$, for isobar collision at $\sqrt{s_{NN}}=$ 200 GeV for 0-5 \% centrality. The symbols are similar as in Fig. \ref{V2sqv4vecRuZr}.}
	\label{V2V3v5vecRuZr}
\end{figure}

In Fig. \ref{V2V3v5vecRuZr}, the  nonlinear correlation coefficient between $V_2 V_3$ and $V_5(p)$ is shown. Here also, the difference between the correlations due to the difference in shape is observed, but it is relatively smaller than for the nonlinear coefficient between $V_2^2$ and $V_4(p)$. The correlation is larger for Zr than Ru, which itself  is  larger than the spherical one. Moreover, it can be seen that the momentum dependence of the decorrelation is significant, similar as for the  $V_2^2-V_4(p)$ correlator.

\begin{figure}[!tbh]
	\vspace{5mm}
	\begin{center}
		\includegraphics[width=0.47\textwidth]{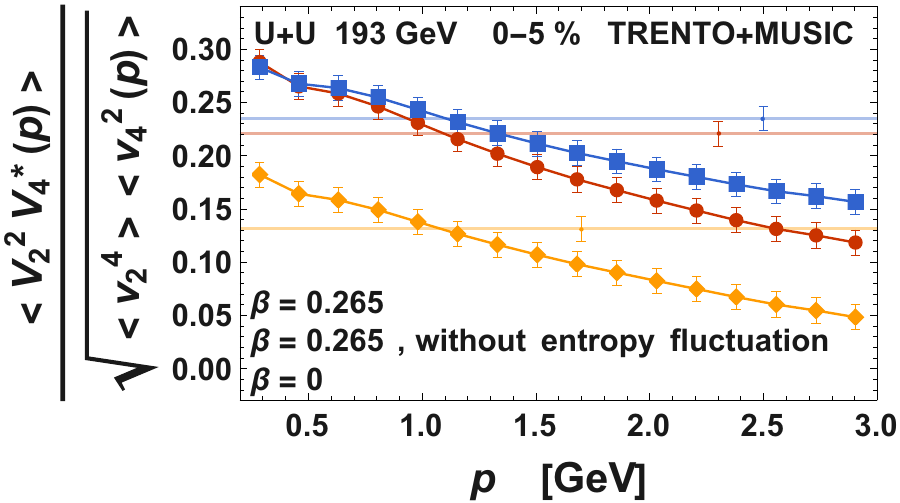} 
	\end{center}
	\caption{Nonlinear flow vector correlation between $V_2^2$ and $V_4(p)$ as a function of transverse momentum $p$, for U + U collision at $\sqrt{s_{NN}}=$ 193 GeV for 0-5 \% centrality. The results with and without fluctuation in the initial entropy deposition are shown by the red dots and blue squares respectively. The orange diamonds represent the results for the corresponding spherical uranium nuclei. The horizontal lines represent the corresponding correlations between the momentum averaged flow vectors. }
	\label{V2sqv4vecUU}
\end{figure}

In Fig. \ref{V2sqv4vecUU} the $V_2^2-V_4(p)$ correlations for central  U+U collisions are shown. As expected, the correlation is larger for collisions of  deformed nuclei  than for spherical ones, following a similar trend for the momentum dependent decorrelation of the elliptic flow (Sec. \ref{sec:UUresults}). The difference between  collisions of deformed and spherical nuclei is  larger for U+U  collisions in comparison to isobar collisions, which is due to a relatively smaller contribution of fluctuations for collisions of larger nuclei.

In the current scope, we also study another two nonlinear mixed-flow correlations, namely $V_1V_3-V_4(p)$ and $V_3^2V_2^*-V_4(p)$ correlations as proposed in \cite{Jia:2022qrq} as the sub-leading non-linear contribution to $V_4$.    

\begin{figure}
 	\vspace{5mm}
 	\begin{center}
 		\includegraphics[width=0.47\textwidth]{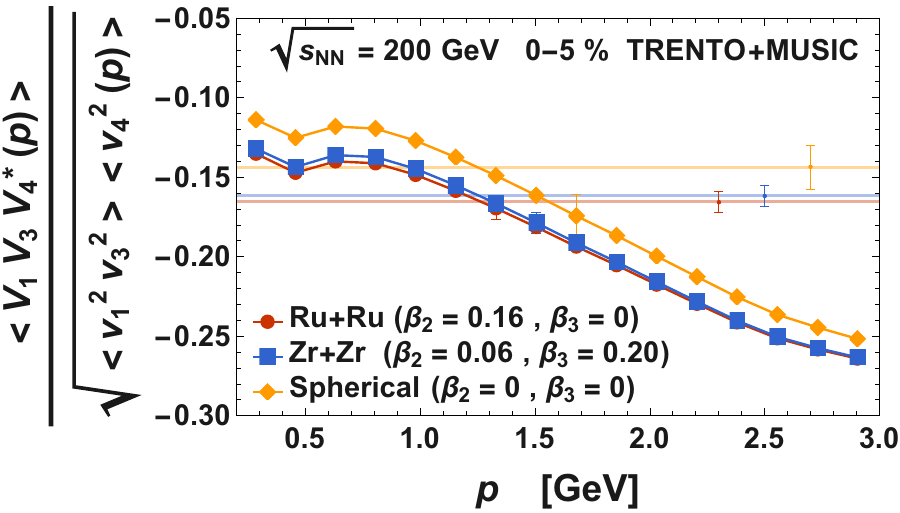} 
 	\end{center}
 	\caption{Sub-leading nonlinear flow vector correlation between $V_1V_3$ and $V_4(p)$ as a function of transverse momentum $p$, for isobar collision at $\sqrt{s_{NN}}=$ 200 GeV for 0-5 \% centrality. The symbols are similar as in Fig.\ref{V2sqv4vecRuZr}.}
 	\label{V1V3v4vecRuZr}
 \end{figure}
\begin{figure}
 	\vspace{5mm}
 	\begin{center}
 		\includegraphics[width=0.47\textwidth]{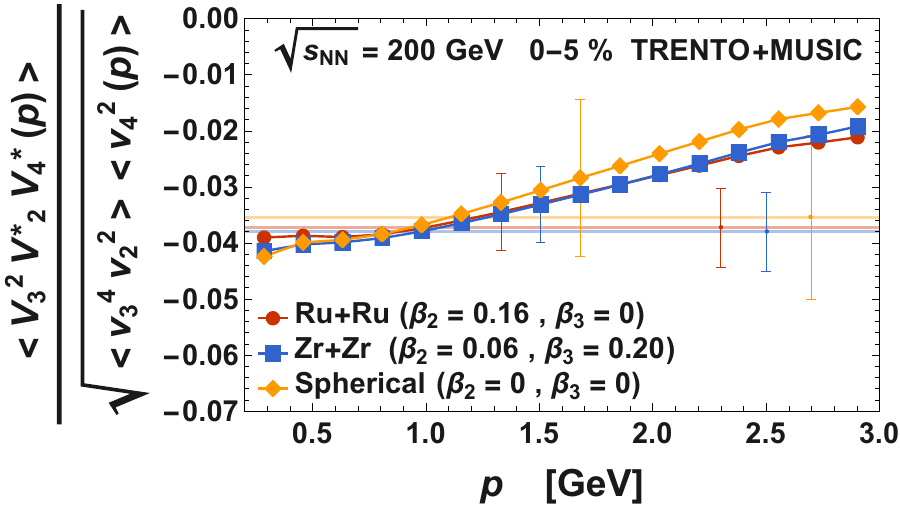} 
 	\end{center}
 	\caption{Sub-leading nonlinear flow vector correlation between $V_3^2V_2^*$ and $V_4(p)$ as a function of transverse momentum $p$, for isobar collision at $\sqrt{s_{NN}}=$ 200 GeV for 0-5 \% centrality. The symbols are similar as in Fig.\ref{V2sqv4vecRuZr}.}
 	\label{V3sqV2sv4vecRuZr}
 \end{figure}
\begin{figure}
	\vspace{5mm}
	\begin{center}
		\includegraphics[width=0.47\textwidth]{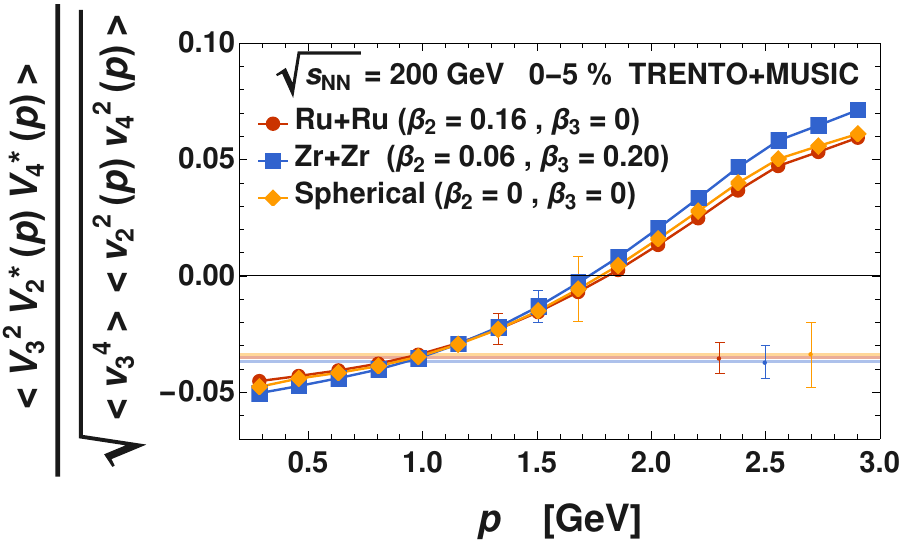} 
	\end{center}
	\caption{Nonlinear flow vector correlation between $V_3^2$ and $V_2(p)V_4(p)$ as a function of transverse momentum $p$, for isobar collision at $\sqrt{s_{NN}}=$ 200 GeV for 0-5 \% centrality. The symbols are similar as in Fig.\ref{V3sqV2sv4vecRuZr}.}
	\label{V3sqv2v4vecRuZr}
\end{figure}

 The usual momentum dependent construction of these correlations would be
 \begin{equation}
 \frac{\langle V_1V_3V_4^*(p)\rangle }{\sqrt{\langle v_1^2v_3^2\rangle \langle v_4^2(p)\rangle}} \ \ \ \text{and} \ \ \	\frac{\langle V_3^2V_2^*V_4^*(p)\rangle }{\sqrt{\langle v_3^4v_2^2\rangle \langle v_4^2(p)\rangle}}  
 	\label{subleadingveccorr}
 \end{equation}  
 The results for the above correlations in the isobar collisions are presented in Fig. \ref{V1V3v4vecRuZr} and \ref{V3sqV2sv4vecRuZr}. From Fig. \ref{V1V3v4vecRuZr}, one can see that the sub-leading non-linear contribution of $V_1V_3$ to $V_4$ is negative and it also shows a significant momentum-dependent decorrelation. On the other hand from Fig. \ref{V3sqV2sv4vecRuZr}, it could be seen that the contribution of $V_3^2V_2^*$ to $V_4$ is very small (an order of magnitude less than the other two contributions) and also negative. 
 However, interestingly it shows a completely opposite trend for the momentum dependence; the correlation coefficient increases with increasing transverse momentum. In this context, we find another interesting correlation: the momentum dependent non-linear correlation between $V_3^2V_2^*$ and $V_4$ could be also constructed as, 
  \begin{equation}
	\frac{\langle V_3^2V_2^*(p)V_4^*(p)\rangle }{\sqrt{\langle v_3^4\rangle \langle v_2^2(p)v_4^2(p)\rangle}}  
 	\label{subleadingveccorr2}
 \end{equation} 
the result of which is presented in Fig. \ref{V3sqv2v4vecRuZr}. The correlation shows a similar behavior as Fig. \ref{V3sqV2sv4vecRuZr}, its negative and increases with momentum with much stronger momentum dependence. Moreover, it could be seen that the correlation changes sign when going from low to high momentum. Such behavior is very peculiar and interesting in the correlation study.

\section{Centrality comparisons and ratio results}

\label{sec:ratioandcent}

\begin{figure}
	\vspace{5mm}
	\begin{center}
		\includegraphics[width=0.45\textwidth]{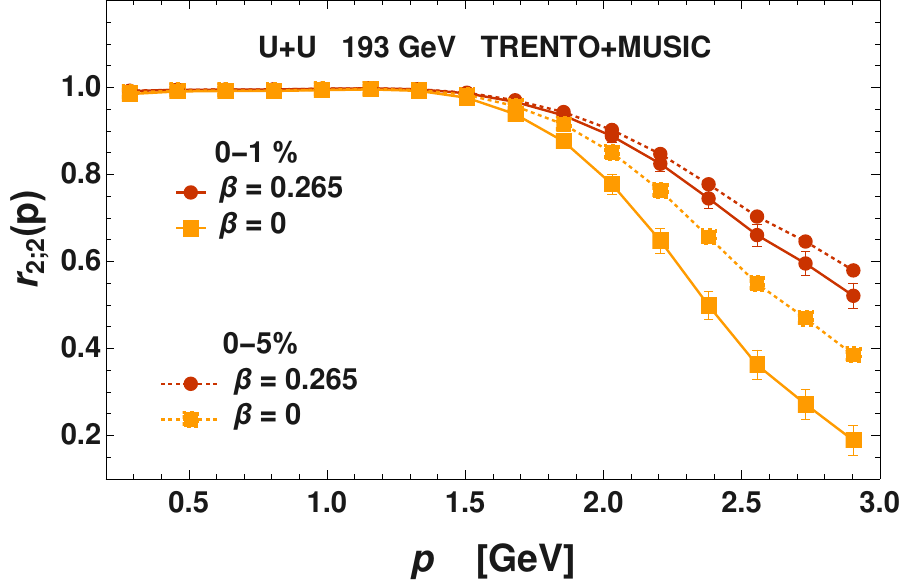} 
	\end{center}
	\caption{Centrality comparison for the $V_2^2-V_4(p)$ correlation as a function of transverse momentum $p$, for U + U collision at $\sqrt{s_{NN}}=$ 193 GeV for 0-5 \% centrality. The deformed and the spherical nuclei collision are denoted by the red dots and orange squares respectively. The solid and the dashed lines represent 0-1 \% and 0-5 \% centrality respectively. }
	\label{centcompV2sqv2sqvecUU}
\end{figure}

In this section, we compare the results    for ultra-central  (0-1 \%) and central (0-5\%) collisions.   When restricting the event sample to ultracentral collisions, the relative importance of the effects of nuclear deformation on initial eccentricities is expected to be larger.
\begin{figure}
	\vspace{5mm}
	\begin{center}
		\includegraphics[width=0.47\textwidth]{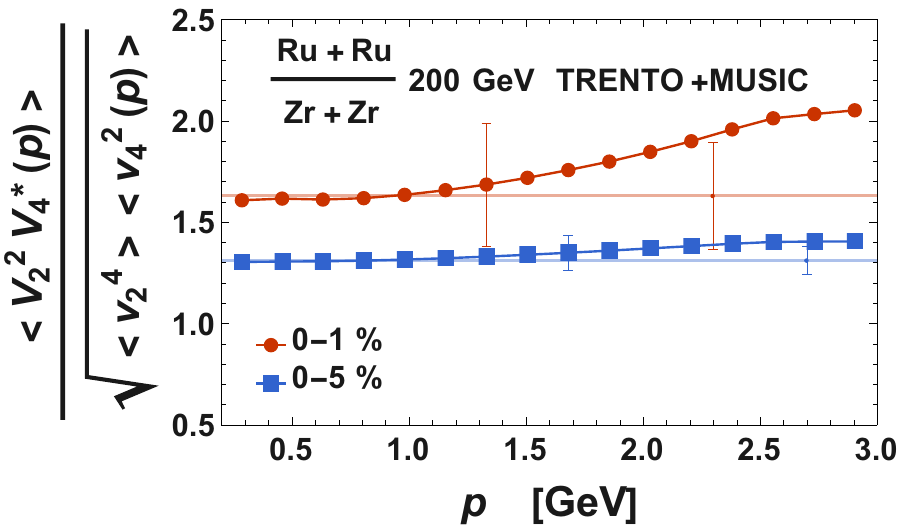} 
	\end{center}
	\caption{The ratio of the $V_2^2-V_4(p)$ correlation between Ru to Zr as a function of transverse momentum $p$, for isobar collision at $\sqrt{s_{NN}}=$ 200 GeV. The results for 0-1\% and 0-5\%  are shown by the red dots and blue squares respectively.}
	\label{ratioV2sqv4vecRuZr}
\end{figure}
In Fig \ref{centcompV2sqv2sqvecUU}, the factorization-breaking coefficient $r_{2;2}(p)$ for U+U collision has been plotted for two centralities, 0-1\% (solid) and 0-5 \%(dashed). The correlation coefficient becomes smaller for ultracentral collisions, which reflects the fact that the eccentricity without deformation is smaller for $0$-$1$\% centrality. On the other hand, the relative difference between the deformed and the spherical case is larger for the 0-1\% than for the 0-5\% event sample.

\begin{figure}
	\vspace{5mm}
	\begin{center}
		\includegraphics[width=0.47\textwidth]{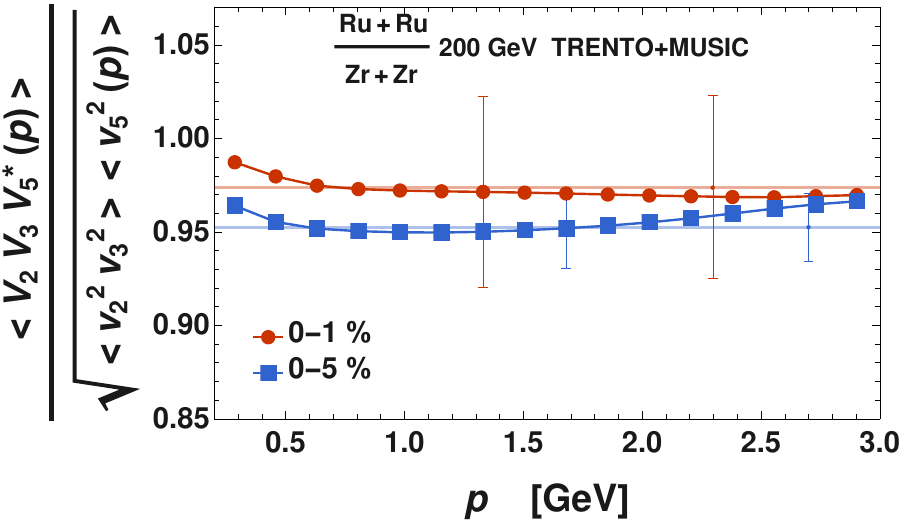} 
	\end{center}
	\caption{The ratio of the $V_2V_3-V_5(p)$ correlation between Ru to Zr as a function of transverse momentum $p$, for isobar collision at $\sqrt{s_{NN}}=$ 200 GeV. The results for 0-1\% and 0-5\%  are shown by the red dots and blue squares respectively.}
	\label{ratioV2V3v5vecRuZr}
\end{figure}

We also study the ratio of the observables between Ru and Zr, which could be measured in experiment at RHIC, as studied by the STAR collaboration \cite{STAR:2021mii}. Any deviation of such ratios from 1 is a consequence of the different deformation  of the isobars. The ratios of the nonlinear correlations $V_2^2-V_4$ and $V_2V_3-V_5$ for the isobar collision were studied recently in \cite{Jia:2022qrq}. We study similar ratios for the momentum dependent nonlinear correlations \ref{mixedflowcorr}. In Fig. \ref{ratioV2sqv4vecRuZr}, the ratio of $V_2^2-V_4(p)$ between Ru to Zr are presented for the two centralities 0-1 \% and 0-5 \%. In both cases, the ratio is larger than 1, which is a clear indication of the increase of the deformation effect  for the  0-1 \% centrality.  Moreover, for 0-1 \% centrality  this ratio is significantly increasing with momentum reaching a value close to 2 at very high transverse
momenta. In Fig. \ref{ratioV2V3v5vecRuZr}, the ratio results for $V_2V_3-V_5(p)$ are shown for the isobar collision. The ratios are consistent with 1 for 0-1 \% centrality but smaller than 1 for 0-5 \% centrality, as expected from Fig. \ref{V2V3v5vecRuZr}. Also we find our baselines of these ratio plots are consistent with the results of \cite{Jia:2022qrq}.

\section{Summary and Outlook}

In this paper, we  look into the momentum dependent flow correlations for collisions of  deformed nuclei  at RHIC energies. We  study how the nuclear deformations affect these correlation coefficients between different flow harmonics.

For   U+U collisions, the deformed nuclear structure of uranium nuclei significantly alters the transverse momentum dependent elliptic  flow  factorization-breaking coefficients when compared to the corresponding spherical case.
A significant  difference is  visible for both flow magnitude and flow
angle decorrelations. Though the difference is very small for isobar collisions.
A qualitatively significant difference is visible for the factorization-breaking coefficient
for the triangular flow, when comparing collisions of nuclei with or without octupole  deformation.

For the isobar collisions of Ru+Ru and Zr+Zr, despite their similarity in the mass number, there is a  difference in the  momentum dependent mixed-flow correlations e.g. $V_2^2-V_4(p)$ and $V_2V_3-V_5(p)$, which arises solely due to the  difference in shape  between the two isobars. The difference is distinctly visible in the ratio results of $V_2^2-V_4(p)$ between Ru to Zr , where the ratio is quite larger than 1 and also varies with the transverse momentum.
We confirm that  the effect of the nuclear deformation on flow correlations strongly depends on the centrality of the collision. A significant difference is observed in the factorization-breaking coefficients at  higher transverse momenta between central ( 0-5\%) and ultra-central ( 0-1\%) collisions.

Moreover, we also study the sub-leading non-linear correlations to $V_4$, namely $V_1V_3-V_4(p)$  and $V_3^2V_2^*-V_4(p)$ correlations, which we find relatively smaller and negative. We also find an unusual momentum dependence for $V_3^2V_2^*V_4(p)$, where the correlation coefficient increases with increasing transverse momentum, although the effect is small. Our model results could be tested and verified in  heavy-ion experiments. Such observables, if tested in experiments, would put new constraints on the existing models and could open another path for the  study  of nuclear structure effects in  relativistic heavy-ion collision.
 
\section*{Acknowledgments}
This research was supported by the AGH University of Science and
Technology and  by the  Polish National Science Centre grant: 2019/35/O/ST2/00357.

\bibliography{ref.bib}

\end{document}